\documentclass[aps,amssymb,amsmath, twocolumn, pra, superscriptaddress]{revtex4}
\usepackage{graphicx}
\usepackage{epsfig}
\usepackage{amsmath}
\usepackage{bm}
\usepackage{bbm}
\usepackage{ulem}
\usepackage[colorlinks, citecolor= magenta]{hyperref}
\usepackage{color}
\usepackage{soul}
\usepackage{centernot}
\usepackage[up]{subfigure}

\usepackage{epstopdf}

\newcommand{\be}{\begin{equation}}
\newcommand{\ee}{\end{equation}}
\newcommand{\bc}{\begin{center}}
\newcommand{\ec}{\end{center}}
\newcommand{\bea}{\begin{eqnarray}}
\newcommand{\eea}{\end{eqnarray}}
\newcommand{\ba}{\begin{array}}
\newcommand{\ea}{\end{array}}

\def\bra#1{\mathinner{\langle{#1}|}}
\def\ket#1{\mathinner{|{#1}\rangle}}

\usepackage{booktabs}

\usepackage{amsthm}
\begin{document}
\title{On witnessing arbitrary bipartite entanglement in a measurement device independent way}
\author{Arindam Mallick}
\email{marindam@imsc.res.in}
\author{Sibasish Ghosh}
\email{sibasish@imsc.res.in}
\affiliation{Optics and Quantum Information Group, The Institute of Mathematical
Sciences, HBNI, C. I. T. Campus, Taramani, Chennai 600113, India}
%================================================

\begin{abstract}
 Experimental detection of entanglement of an arbitrary state of a
 given bipartite system is crucial for 
exploring many areas of quantum information. But such a detection should be made in a device independent way if the 
preparation process of the state is considered to be faithful,  in order to avoid detection of a separable state as entangled one.
The recently developed scheme of detecting bipartite entanglement 
in a measurement device independent way [Phys. Rev. Lett {\bf 110}, 060405 (2013)]
does require information about the state. Here by using 
Auguisiak et al.'s universal entanglement witness scheme for two-qubit states [Phys. Rev. A {\bf 77}, 030301 (2008)], 
we provide a universal detection scheme for two-qubit states in a measurement device independent way. 
We provide a set of universal 
witness operators for detecting  NPT-ness(negative under partial transpose) of  
states in a measurement device independent way.
We conjecture that no such universal entanglement witness 
exists for PPT(positive under partial transpose) entangled states.  We also analyze
the robustness of some of the experimental schemes of detecting entanglement  in a measurement device independent way under the influence of 
noise in the inputs (from the referee) as well as in the  measurement operator as envisazed in ref.  [Phys. Rev. Lett {\bf 110}, 060405 (2013)].
\end{abstract}
\maketitle
\section{Introduction:}
Entanglement is shown to be a resource for quantum information proccesing, like  quantum cryptography \cite{crypto},
telepotation \cite{tele}, quantum metrology \cite{metrol}, channel capacity \cite{capa, dense}.
Moreover, it helps to speed up computation sometimes \cite{computa} 
over the existing classical algorithms. Also,  understanding entanglement is neccessary to reveal various non-classical nature of the physical world.  
Entanglement in quantum theory correlates two parties in such a way that at the individual levels,  they lose their independent state descriptions. 
Mathematically, two parties $\mathcal{A} $ and $\mathcal{ B} $ are separable iff their joint state $\rho $ can be expressed as
$ \sum_j \alpha_j ~ \mathcal{\rho_A}^j \otimes \mathcal{\rho_B}^j $ acting on the Hilbert space
$ \mathcal{H_{AB}} = \mathcal{H_A \otimes H_B} $, where
$\sum_j \alpha_j = 1 $ and $ 0\leq \alpha_j \leq 1$  $ \forall j $ ,  $\mathcal{\rho_A}^j $  and  $ \mathcal{\rho_B}^j $ are density 
matrices acting on  $ \mathcal{H_A} $ and $ \mathcal{H_B} $ respectively.
If the joint state can't be written in the above 
form,  we call them entangled  \cite{Asher}.

Deciding whether a unknown state of a given bipartite quantum system is entangled or separable,
remains as a challenging problem \cite{rhoro,loannou,loannou2} right from the initial stage of quantum information theory.
People tried to see the presence of entanglement in a state experimentally through some Bell-inequality violation,
but there are entangled 
states which do not violate any such inequality  \cite{Horo}.
Given any entangled state $ \rho_\mathcal{AB}$, it is (in principle) possible to find out an entanglement witness observable  $W_{\rho}$,
whose measurement can distinguish the given entangled states from all possible separable states of the system, and the measurement is, in principle,
 implementable experimentally, even using local measurement settings, (possibly) supported by classical communication.
%what mathematical theorem say this ?
It may be noted that an entanglement witness is a hermitian operator $ W : \mathcal{H_A}\otimes \mathcal{H_B} \rightarrow \mathcal{H_A}\otimes \mathcal{H_B}$
such that there exists atleast one entangled state $\rho_{\mathcal{AB}}$ on $\mathcal{H_A}\otimes \mathcal{H_B},$ for which $\text{Tr}[W\rho_{\mathcal{AB}}]$ $<$ 0 
while $\text{Tr}[W\sigma_{\mathcal{AB}}$] $\geqslant$ 0 for all separable density matrices $\sigma_{\mathcal{AB}}$ on $\mathcal{H_A}\otimes \mathcal{H_B}$. 
So W witnesses the entanglement in the state $\rho_{\mathcal{AB}}$. 
But  $\text{Tr}[ W\sigma_{\mathcal{AB}}] \geqslant 0 $  does not necessarily imply $\sigma_{\mathcal{AB}} $ is separable
\cite{Tura}. In general, $W_{\rho}$ depends on the the form of entangled state $\rho_{\mathcal{AB}}$, and hence it does not have a universal character. 
Thus a state-independent witness operator is a desired one for most practical 
purposes.

The conventional entanglement detection strategies are 
based on the local measurement on each subsystem  $\mathcal{A}$ and $\mathcal{B} $.
In this case,  perfectness of both the local measuring apparatuses is necessary. 
In fact, if some measurement results get lost or some over counting occurs in the measurement,
it may lead to an erroneous conclusion about the entanglement of the state $\rho_{\mathcal{AB}}$. 
In  ref. \cite{10}, time shift attack has been
used experimentally which causes identification of a separable state as an entangled one.
Such a scenario can be avoided if one can witness entanglement in a measurement device independent way.

Recently, Branciard et al. \cite{Gisin-mdiew} demonstrated implementation of any
entanglement witness operator  
in a measurement-device-independent way using Buscemi's work on semi-quantum nonlocal game \cite{Buscemi}.
This scheme of Branciard et al. \cite{Gisin-mdiew} guaranteed that no separable state will be detected as an entangled state, irrespective of the kind of noise 
effects present during the measurement \cite{Gisin-mdiew}. But their scheme depends on the form of the shared entangled state. Hence their witness 
operator lacks universal character, that is, it is unable to detect entanglement in a unknown state of the bipartite system.

Augusiak et al. \cite{augusiak} showed the existence of a universal entanglement 
witness operator that can detect entanglement in any two-qubit state, depending on the Peres-Horodecki PPT criteria \cite{Asher, decki}.
But this does not address measurement device independent implementation of the witness operator, and moreover it is restricted to two-qubit case only.    

 In this paper, we claim that this two-qubit universal entanglement witness operator of Augusiak et al. \cite{augusiak}
 can be implemented in a measurement device independent 
 way.  We then generalize this idea for higher dimensional bipartite NPT entangled states and find out
 a set of finite number of universal NPT witnesses for any {\it given} bipartite system.
 We conjecture that no universal entanglement witness operator can exist to detect entanglement in a PPT entangled state.
 We also point out that, in the aforesaid scheme of Branciard et al.'s measurement device independent entanglement
 witness scheme \cite{Gisin-mdiew},  if some general 
 type of noise is added with the inputs or with the measurement operator, the referee will not face any
 problem in witnessing entanglement provided he knows the 
 character of the noise (here we will provide analysis for a particular type of noise and general treatment in appendix \ref{secfive}).  
 In case the referee doesn't know, we show how robust the witnessing operation can be dealt with.

Our article is organized as follows. In section \ref{firsts}, we describe measurement device independent (MDI) scheme of entanglement witness (EW).
Section \ref{sectwo} is devoted to MDI implementation of the two-qubit universal EW of Auguisiak et al. \cite{augusiak}. \ref{secthree} describes universal witness for NPT-ness
of two-qudit states. Section \ref{secfour}, describes the conjecture about the impossibility of the existence of universal witness operator for 
bipartite PPT-entangled states. In section \ref{secfive}, we analyse the possible noise effects in measurement and inputs. We draw the conclusion of our work 
in section \ref{secsix}.
 
 \section{Measurement Device Independent(MDI) Scheme of Entanglement Witness (EW) :}\label{firsts}
 
 This scheme starts as a cooperative game played between two parties Alice($ \mathcal{A}$) and Bob($ \mathcal{B}$),  who receive quantum
states \{$\tau_s \in \mathcal{H_{A_O}}$\} and \{$\omega_t \in \mathcal{H_{B_O}}$\} respectively as inputs from a referee, who then ask them to 
produce outputs $ a $ and $ b $ respectively.  Here dim($\mathcal{H_{A_O}}$) = dim($\mathcal{H_A}$)= $ d_{\mathcal{A}} $ and 
dim($\mathcal{H_{B_O}}$) = dim($\mathcal{H_B}$) = $ d_{\mathcal{B}} $  and for our analysis, we consider $ d_{\mathcal{A}} = d_{\mathcal{B}} = d $. 
During the game, Alice and Bob aren't allowed to communicate anyway,
but before starting the game they have to decide their strategies
where they may share apriori a bipartite state $\rho_{\mathcal{AB}}$. If the state $\rho_{\mathcal{AB}}$ is entangled 
state they can always get an average pay-off which is more 
than the case when $ \rho_{\mathcal{AB}} $ is any separable state.  So, from the average pay-off values, the referee can conclude
whether the corresponding shared state was entangled 
or separable ( by knowing the maximum average payoff for all possible separable states).  
To produce outcomes of the game, each party has to perform local joint projective measurement on their individual inputs and 
their individual subsystems of the shared state, see Figure 1 for details.
\begin{figure}[center]\label{fig1}
 \includegraphics[scale = 0.37]{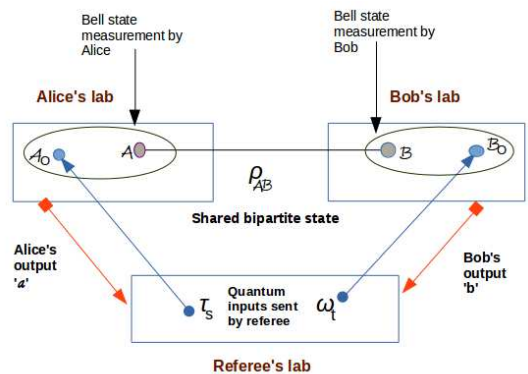}
 \caption{(colour online) Alice and Bob performs Bell state measurement on the states
  of the joint systems $\mathcal{A_O A}$ and $\mathcal{B B_O}$ respectively.  Blue arrows are 
  lossless quantum channels for sending input states to Alice and Bob.  Red arrows correspond to the 
  lossless classical channels for receiving the outputs. Green ellipces indicate joint Bell state measurement 
  performed by the players.}
 \end{figure}
 
The authors of \cite{Gisin-mdiew} described the aforesaid projective measurement in the maximally entangled state 
$$\ket{\Phi}_{\mathcal{A_O A}} =\ket{\Phi}_{\mathcal{B B_O}} = \frac{1}{\sqrt{d}} \sum\limits_{j = 0}^{d-1}  \ket{jj}  \in  \mathbb{C}^d \otimes \mathbb{C}^d ,$$ 
 with probability of output $(a,b) = (1, 1)$ for given quantum input pair ($\tau_s,\omega_t $) being $ P_{\rho_{\mathcal{AB}}}(1,1|s,t) =
 \text{Tr}\Big((\ket{\Phi}_{\mathcal{A_O A}}\bra{\Phi}\otimes\ket{\Phi}_{\mathcal{B B_O}}\bra{\Phi}) 
   \big( \tau_s\otimes\rho_{\mathcal{AB}}\otimes\omega_t \big) \Big), $  and showed that if  
the average payoff function, $\Pi(\rho_{\mathcal{AB}}) = \sum_{s,t} \tilde{\beta}_{s,t} P_{\rho_{\mathcal{AB}}}(1,1|s,t) $ then
\begin{align}
  I(\rho_{\mathcal{AB}})  := \max \big\{ \Pi( \sigma_{\mathcal{AB}}) \big\} -\Pi(\rho_{\mathcal{AB}}) ~~~~~  \nonumber\\
  =  \max\limits_{\sigma_{\mathcal{AB}}} \Big\{ \sum_{s,t} \tilde{\beta}_{s,t} P_{\sigma_{\mathcal{AB}}}(1,1|s,t) \Big\}
  - \sum_{s,t} \tilde{\beta}_{s,t} P_{\rho_{\mathcal{AB}}}(1,1|s,t) \nonumber\\
 = \sum_{s,t} \beta_{s,t} P_{\rho_{\mathcal{AB}}}(1,1|s,t).~~~~~~~~~~~~~~~~~~~~~ 
\end{align}
where, the maximization is performed over all possible separable state $\sigma_{\mathcal{AB}}$ of the bipartite system.
The outcome (1,1) is nothing but the successful projection 
on the state $ \tau_s\otimes\rho_{\mathcal{AB}}\otimes\omega_t  $ by the tensor product of the operator 
$\ket{\Phi}_{\mathcal{A_O A}} \bra{\Phi} \otimes \ket{\Phi}_{\mathcal{B B_O}} \bra{\Phi} $ 
and $\beta_{s,t}$ is pay-off value corresponding to the input state $( \tau_s, \omega_t)$.

If we want to use  $ I(\rho_{\mathcal{AB}})$ as a signature of entanglement witness,
it has to be proportional to $\text{Tr}[W\rho_{\mathcal{AB}}$], and
the witness operator $ W $ needs to be expressed as a linear combination 
of the tensor product of the transposed input states,
\begin{align} \label{beta}
W = \sum_{s,t} \beta_{s,t} \tau_s^{T}\otimes\omega_t^{T},\end{align}

which,  in principle, is possible in this case as the inputs $\tau_s$ (to Alice)
and $\omega_t $ (to Bob)-supplied by the referee- are sufficient enough to span the 
respective vector space of witness operators on $ \mathcal{H_A} \otimes \mathcal{H_B}$.
Then the referee will determine the values of  $\beta_{s,t}$
accordingly prior to the game, such that he can identify whether the shared state is entangled from the sign of $I(\rho_{\mathcal{AB}}$).
Indeed this can be done as
$W$ is hermitian and transposed input density matrices are also hermitian.
The referee just have to choose the input states $\tau_s$ and $\omega_t$ in order to form a complete set of operators 
\{ $\tau_s^{T}\otimes\omega_t^{T} $ \}  to express $ W $ as a linear combination of these operators.

 In this scenario, accuracy in preparation of the input states must be trusted.
 In ref. \cite{Gisin-mdiew}, it is proved that no noise introduced in the local measurement
 operator can lower the bound $ I(\rho_{\mathcal{AB}})\geq 0$  for separable $\rho_{\mathcal{AB}} $, 
 so, in this scheme, separable state will never be identified 
 as an entangled one.
 In this sense the scheme of ref. \cite{Gisin-mdiew} is measurement device independent.

 \subsection{ \large Detecting entanglement in unknown two-qubit state :} 

 In the case of witnessing in a unknown two-qubit state \cite{augusiak},  instead of sharing single copy of the state,  the players need to 
share four identical copies of the state. This is so because the signature of det($\rho_{\mathcal{AB}}^{T_{\mathcal{B}}} $)
= product of eigenvalues of   $\rho_{\mathcal{AB}}^{T_{\mathcal{B}}} $ = $ \text{Tr}\big[ W \rho_{\mathcal{AB}}^{\otimes 4} \big] = $ 
$\lambda_1 \lambda_2 \lambda_3 \lambda_4$, completely determines whether the two-qubit state $ \rho_\mathcal{AB}$ is entangled or separable.
Here $\lambda_i $'s are eigenvalues of $\rho^{T_{\mathcal{B}}}_{\mathcal{AB}}$ 
and $W$ is a hermitian operator acting on $ \mathcal{H}_\mathcal{A}^{\otimes 4} \otimes \mathcal{H}_\mathcal{B}^{\otimes 4} $ 
(with dim$\mathcal{H_A}$ = dim$\mathcal{H_B}$ = 2).
From the Newton-Girard formula \cite{Mead}, we have : $ \lambda_1 \lambda_2 \lambda_3 \lambda_4 = $
\begin{align}\frac{1}{24}\bigg(1-6\sum_{i=1}^4\lambda_i^4 + 8\sum_{i=1}^4\lambda_i^3
+ 3\Big(\sum_{i=1}^4 \lambda_i^2\Big)^2 - 6\sum_{i=1}^4 \lambda_i^2 \bigg) \nonumber \end{align}

To formulate this in terms of an operator, we first use the swap operator
$V^{(k)},$ defined as
 \begin{align} V^{(k)} \big( \ket{\phi_1}_{\mathcal{A}_1 \mathcal{B}_1 } 
\otimes \ket{\phi_2}_{\mathcal{A}_2 \mathcal{B}_2 } \otimes....\ket{\phi_k}_{\mathcal{A}_k \mathcal{B}_k }  \big) \nonumber\\
= \ket{\phi_k}_{\mathcal{A}_1 \mathcal{B}_1 } \otimes \ket{\phi_1}_{\mathcal{A}_2 \mathcal{B}_2 } \otimes....\ket{\phi_{k-1}}_{\mathcal{A}_k \mathcal{B}_k } 
\end{align}
 with the property \text{Tr}($V^{(k)} \rho^{\otimes k}$) = \text{Tr}($\rho^k$) \cite{augusiak,Pawel,ekert} 
 for any hermitian matrix $\rho$ and $ V^{(k)} $ is not, in general,  a hermitian operator except for $k = 2$.
 So to make it an observable in place of 
 $V^{(k)},$  $ \frac{1}{2}(V^{(k)} + V^{(k)\dagger}) $  will be used for farther analysis. 
 As the swap operator $V^{(k)}$ is a real matrix with respect to a fixed basis,
 used for the whole analysis, 
\begin{align} \text{Tr}((\rho_{\mathcal{AB}}^{T_{\mathcal{B}}})^k)=
 \text{Tr}(V^{(k)} (\rho_{\mathcal{AB}}^{T_{\mathcal{B}}})^{\otimes k}) \nonumber\\
     = \frac{1}{2} \text{Tr}\Big((V^{(k)} + V^{(k)\dagger})^{T_{\mathcal{B}}}\rho_{\mathcal{AB}}^{ \otimes k} \Big) \nonumber\\     = 
 \frac{1}{2} \text{Tr} \Big( ~ \Big[~ \tilde{ V}^{(k)} \otimes \tilde {V}^{(k)T}
 + \tilde{ V}^{(k)T} \otimes \tilde {V}^{(k)} ~ \Big] ~ \rho_{\mathcal{AB}}^{\otimes k} \Big), \nonumber\\
 \text{ where}~ \tilde{V}^{(k)} : \mathcal{H}_j^{\otimes k} \rightarrow \mathcal{H}_j^{\otimes k}  \nonumber \end{align} 
 is given by : 
 \begin{align}
  \tilde{V}^{(k)} \big( \ket{\phi_1}_{j_1} \otimes  \ket{\phi_2}_{j_2} \otimes ...... \otimes \ket{\phi_k}_{j_k}   \big) \nonumber\\
  = \ket{\phi_k}_{j_1} \otimes  \ket{\phi_1}_{j_2} \otimes ...... \otimes \ket{\phi_{(k-1)}}_{j_k} ~~\text{for}~ j = \mathcal{A}, \mathcal{B}.
 \end{align}

 Then, hermitian operator witnessing entanglement in arbitrary two-qubit state, is given by,
 \begin{widetext}
 \begin{align}\label{univwit}
   W^{\text{univ}} = \frac{1}{24}\mathbb{I}_{256 \times 256} - \frac{1}{8}\bigg(\tilde{V}^{(4)}\otimes\tilde{V}^{(4)T}
  + \tilde{V}^{(4)T}\otimes\tilde{V}^{(4)}\bigg)% \nonumber \\
  +\frac{1}{6}\mathbb{I}_{4\times 4} \otimes\bigg(\tilde{V}^{(3)}\otimes\tilde{V}^{(3)T} 
 + \tilde{V}^{(3)T}\otimes\tilde{V}^{(3)} \bigg)  \nonumber\\
 +  \frac{1}{8} V^{(2)}\otimes V^{(2)}-  
 \frac{1}{4} \mathbb{I}_{16 \times 16} \otimes V^{(2)}, ~~~~~~~~~~~~~~~~~~~~~~~~~~~~~~~~~~~~~~~~
 \end{align}
 \end{widetext}
 \vspace{1em}

 and our $I(\rho_{\mathcal{AB}}$) turns out to be proportional to \text{Tr}($W^{\text{univ}} \rho_{\mathcal{AB}}^{\otimes 4 }$) = 
det( $\rho_{\mathcal{AB}}^{T_{\mathcal{B}}}$),  see the next section \ref{sectwo}.
 
  The above form of witness operator doesn't depend on the type of two-qubit shared state,
  so that entanglement in an unknown 
  state can be detected. In this sense it's universal entanglement witness operator.

 \section{MDI Implementation of Two-qubit Universal EW :}
 \label{sectwo}
  Recent work by Bartkiewicz et al. in  \cite{Bartkiewicz}  demonstrated the implementation of the 
  aforesaid universal entanglement witness operator (\ref{univwit}) for two-qubit case,  using photon polarized states.
  But that has not been done in a measurement device independent way.
  
   Authors of ref. \cite{Yuan} and \cite{Verbanis}, address entanglement detection of arbitrary two-qubit states in MDI way, but their approach 
   focus on optimization of entanglement witness operator instead of the question of the universality of that, as we discuss throughout this article.
 
 The game of Fig 1, starts with two inputs (one for Alice and one for Bob) and single 
 copy of a shared state, while our universal measurement scheme for witnessing entanglement needs
 four identical copies of shared state $ \rho_\mathcal{AB}$ at each run of the experiment and the witness operator $W^{\text{univ}} $ is a 256$\times$256
 dimensional matrix. Thus, the input Hilbert space is $ [\mathbb{C}^2 ]^{\otimes 4} \otimes [\mathbb{C}^2 ]^{\otimes 4} = \mathbb{C}^{16} \otimes \mathbb{C}^{16} $
 instead of 
 $ \mathbb{C}^2 \otimes \mathbb{C}^2 $. As a result of that, both $ \tau_s $ and  $ \omega_t $ must be density 
 matrices on $  \mathbb{C}^{16}.$
 
 So, the witness operator $W^{\text{univ}} $ of equ. (\ref{univwit}) should be of the form,

 \begin{align}\label{witform} W^{\text{univ}} = \sum_{s,t}\beta_{s,t} (\tau_s^{T}\otimes\omega_t^{T}), \end{align}
  where, the input states $ \tau_s $ and $ \omega_t $ are expandable in appropriate Gell-mann matrix basis such that

  \begin{align} \tau_s^T  = \frac{ \vec{\Lambda}.\text{Tr}(\tau_s^T \vec{\Lambda})}{16} = \sum_i \Lambda_i ~ p^{(s)}_i, \nonumber\\
  \omega_t^T  = \frac{\vec{\Lambda}.\text{Tr}(\omega_t^T \vec{\Lambda})}{16} = \sum_j \Lambda_j ~ q^{(t)}_j,\end{align}
  
 where  $\vec{\Lambda}$ is the vector of the  $16 \times 16 $  generalized Gell-Mann matrices  \cite{gellmann},
  whose first component represents identity matrix. Here, $ i, j \in  \{0,1,2,3,...,255\} $ 
  and $\Lambda_i $,  $\Lambda_j $ are $i,  j-$th components of ~~
  $\vec{\Lambda} = \big(~ \mathbb{I},~  \Lambda_1,~ \Lambda_2, ~ \Lambda_3,~ ...,  \Lambda_{254}, ~ \Lambda_{255}~ \big)$ respectively.
  Also $p^{(s)}_i$ and $q^{(t)}_j $ are real coefficients.

  Then from the above relation (\ref{witform}) and using the actual form of the universal witness operator 
  (\ref{univwit}), the referee will calculate the quantities,
  $$\text{Tr}\big(  W^{\text{univ}} ( \Lambda_{i}\otimes\Lambda_{j}) \big) =  (16)^2  \sum_{s,t} \beta_{s,t} ~ p^{(s)}_i ~ q^{(t)}_j, $$ 
  that will give a set of $(256)^2$ number of linear equations for all $i, j,$ from 
  which the referee calculates the coefficients $\beta_{s,t}$ in equ. (\ref{witform}).  Also, instead of using a 
  two-qubit Bell state projector  $  \ket{\phi^+} \bra{\phi^+}  =  \frac{1}{2} \sum_{i,j= 0}^{1} \ket{ii}\bra{jj}  $ 
  or it's local unitary equivalent as a measurement operator,
  each party should use Bell state projector of two sixteen dimensional systems as the measurement 
  operator $ \ket{\Phi^+}\bra{\Phi^+} = \frac{1}{16} \sum_{i,j = 0}^{15} \ket{ii}\bra{jj} $ or it's local unitary equivalent, for each party.
  Then the negativity of
  \begin{widetext}
  \begin{align} 
                          I(\rho_{\mathcal{AB}}) = \text{det}(\rho_{\mathcal{AB}}^{T_{\mathcal{B}}}) 
                          = \text{Tr} \big[ W^{\text{univ}} ( \rho_{\mathcal{A}_1 \mathcal{B}_1} \otimes  
                          \rho_{\mathcal{A}_2 \mathcal{B}_2} \otimes \rho_{\mathcal{A}_3 \mathcal{B}_3} 
                          \otimes \rho_{\mathcal{A}_4 \mathcal{B}_4}  ) \big]  ~~~~~~~~~ ~~~~~~ ~~~~~ ~~ \nonumber\\
                          = \sum_{s,t} \beta_{s,t} \text{Tr} \Big[ \Big( \ket{\Phi^+}_{\mathcal{A}_0 :\mathcal{A}_1 \mathcal{A}_2 \mathcal{A}_3 \mathcal{A}_4} \bra{\Phi^+} 
                          \otimes \ket{\Phi^+}_{\mathcal{B}_0 :\mathcal{B}_1 \mathcal{B}_2 \mathcal{B}_3 \mathcal{B}_4} \bra{\Phi^+} \Big) \Big( \tau_s^{ (\mathcal{A}_0)} 
                          \otimes \rho_{\mathcal{A}_1 \mathcal{B}_1} \otimes  
                          \rho_{\mathcal{A}_2 \mathcal{B}_2} \otimes \rho_{\mathcal{A}_3 \mathcal{B}_3} 
                          \otimes \rho_{\mathcal{A}_4 \mathcal{B}_4}  \otimes \omega_t^{ (\mathcal{B}_0)} \Big)  \Big]
                         \end{align}
 \end{widetext}
  confirms the presence of entanglement of the shared state.  
  
  Here if we choose to work in the computational basis $\{ \ket{0}, \ket{1} \},$ any 
  pure state of a single party $\mathcal{A}$, can be written as,
  
  $ \ket{\psi_\mathcal{A}} = \big( \alpha \ket{0}_1 + \beta_1 \ket{1} \big)_{\mathcal{A}_1} \otimes \big( \alpha_2 \ket{0} + \beta_2 \ket{1} \big)_{\mathcal{A}_2}$ \\
 $ \otimes \big(  \alpha_3 \ket{0} + \beta_3 \ket{1}  \big)_{\mathcal{A}_3} \otimes  \big( \alpha_4 \ket{0} + \beta_4 \ket{1} \big)_{\mathcal{A}_4},$
 
     such that the set $\{\ket{\psi_\mathcal{A}}, \ket{\psi_\mathcal{B}}   \}$ spans the support of $\rho_{\mathcal{AB}}^{\otimes 4 }$.
  
 The form of the swap operators has to be found out by their action, 
 
 \begin{align}
  \tilde{V}^{(4)}   \ket{\psi_\mathcal{A}} =   \big( \alpha_4 \ket{0} + \beta_4 \ket{1} \big)_{\mathcal{A}_1}  \otimes
  \big( \alpha_1 \ket{0} + \beta_1 \ket{1} \big)_{\mathcal{A}_2}  \nonumber\\
  \otimes \big( \alpha_2 \ket{0} + \beta_2 \ket{1} \big)_{\mathcal{A}_3}  \otimes
 \big(  \alpha_3 \ket{0} + \beta_3 \ket{1}  \big)_{\mathcal{A}_4}
 \end{align}

 Therefore,
 \begin{align}
   \tilde{V}^{(4)} = \sum_{i,j,k,l = 0}^1 \ket{l, i, j, k}\bra{i, j, k, l} 
  ~~\text{acts on} ~ \mathcal{H_A}^{\otimes 4} ~~ 
 \end{align}

 \begin{align}
  \tilde{V}^{(3)} \otimes  \mathbb{I}   \ket{\psi_\mathcal{A}} =   \big( \alpha_3 \ket{0} + \beta_3 \ket{1} \big)_{\mathcal{A}_1}  \otimes
  \big( \alpha_1 \ket{0} + \beta_1 \ket{1} \big)_{\mathcal{A}_2}   \nonumber\\
 \otimes \big( \alpha_2 \ket{0} + \beta_2 \ket{1} \big)_{\mathcal{A}_3} \otimes
 \big(  \alpha_4 \ket{0} + \beta_4 \ket{1}  \big)_{\mathcal{A}_4}
 \end{align}
  Therefore,
 \begin{align}
   \tilde{V}^{(3)} = \sum_{i,j,k = 0}^1 \ket{k, i, j}\bra{i, j, k}
 ~~ \text{acts on} ~~~ \mathcal{H_A}^{\otimes 3} ~~ 
 \end{align}
  Similar expressions are valid for the operaors act on $\mathcal{H_B}^{\otimes n}~\text{where,}~ n = 3, 4.$
  
  A general support vector in $\mathcal{H_{AB}}^{\otimes 4}$ can be written as,
 \begin{align}
  \ket{\psi_{\mathcal{AB}}} = \big( \alpha_1 \ket{0} + \beta_1 \ket{1} \big)_{\mathcal{A}_1} 
 \otimes \big( \alpha'_1 \ket{0} + \beta'_1 \ket{1} \big)_{\mathcal{B}_1}  \nonumber\\
 \otimes \big( \alpha_2 \ket{0} + \beta_2 \ket{1} \big)_{\mathcal{A}_2} 
 \otimes  \big( \alpha'_2 \ket{0} + \beta'_2 \ket{1} \big)_{\mathcal{B}_2}  \nonumber\\
 \otimes \big( \alpha_3 \ket{0} + \beta_3 \ket{1} \big)_{\mathcal{A}_3} 
 \otimes  \big( \alpha'_3 \ket{0} + \beta'_3 \ket{1} \big)_{\mathcal{B}_3}  \nonumber\\
 \otimes\big(  \alpha_4 \ket{0} + \beta_4 \ket{1}  \big)_{\mathcal{A}_4} 
  \otimes\big(  \alpha'_4 \ket{0} + \beta'_4 \ket{1}  \big)_{\mathcal{B}_4}
 \end{align}
 
 Therefore, 
 \begin{align}
 V^{(2)} \otimes V^{(2)} \ket{\psi_{\mathcal{AB}}} ~~~~~~~~~ ~~~~~~   \nonumber\\
  =   \big( \alpha_2 \ket{0} + \beta_2 \ket{1} \big)_{\mathcal{A}_2} 
  \otimes \big( \alpha'_2 \ket{0} + \beta'_2 \ket{1} \big)_{\mathcal{B}_2}  \nonumber\\
  \otimes \big( \alpha_1 \ket{0} + \beta_1 \ket{1} \big)_{\mathcal{A}_1} 
 \otimes \big( \alpha'_1 \ket{0} + \beta'_1 \ket{1} \big)_{\mathcal{B}_1} \nonumber\\
 \otimes  \big(  \alpha_4 \ket{0} + \beta_4 \ket{1}  \big)_{\mathcal{A}_4} 
  \otimes\big(  \alpha'_4 \ket{0} + \beta'_4 \ket{1}  \big)_{\mathcal{B}_4} \nonumber\\
  \otimes \big( \alpha_3 \ket{0} + \beta_3 \ket{1} \big)_{\mathcal{A}_3} \otimes
   \big( \alpha'_3 \ket{0} + \beta'_3 \ket{1} \big)_{\mathcal{B}_3}
 \end{align}

Thus, \begin{align}
       V^{(2)} =  \sum_{i,j,k,l = 0}^1 \ket{kl, ij}\bra{ij, kl} 
       ~~ \text{acts on} ~~ \mathcal{H_{AB}}^{\otimes 2}
      \end{align}

  \section{\normalsize Universal Witness for NPT-ness Of Two-Qudit states:}
  \label{secthree}
  Any two-qudit density matrix which has at least one negative eigenvalue 
  after partial transposition, is called an NPT state.
  Any bipartite state having NPT is neccessarily entangled.  Now, to determine 
  NPT-ness of a unknown two-qudit state (i.e. to determine whether the state has NPT or PPT),  we need to check negativity of 
  the eigenvalues of $\rho_{\mathcal{AB}}^{T_{\mathcal{B}}}$ \cite{Ekert}.
   Let us look at the characteristic equation for the matrix $\rho_{\mathcal{AB}}^{T_{\mathcal{B}}} :$ 
   \begin{align}\label{charequ}
   \text{det}(\rho_{\mathcal{AB}}^{T_{\mathcal{B}}} - 
   \lambda \mathbb{I}_{d^2 \times d^2}) = 0, ~~~~~ ~~~ ~~~~~~ ~~ \nonumber\\
   \text{which is in the form,}~~ \sum_{i=0}^{d^2} (- 1 )^{d^2 - i} a_{d^2 - i} \lambda^i  = 0, ~~~~~~ ~~~ \end{align}
   where, the coefficients $a_0, a_1, a_2, ...., a_{d^2} $ 
   can be given in terms of the eigenvalues $ \lambda_1,  \lambda_2,  \lambda_3, ... , \lambda_{d^2} $ of $ \rho_\mathcal{AB} $ as follows :
   \begin{align} a_{0} = 1 , a_{1} =  \sum_{i=1}^{d^2} \lambda_i , 
   a_{2} =  \sum_{i,j=1 ;i>j}^{d^2} \lambda_i \lambda_j, \nonumber\\
   a_{3} = \sum_{i,j,k=1;i>j>k}^{d^2} \lambda_i \lambda_j \lambda_k ,
 ......,a_{d^2} =  \prod_{i=1}^{d^2}\lambda_i \nonumber\end{align} 
  
  As $\rho_{\mathcal{AB}}^{T_{\mathcal{B}}} $ is hermitian, 
   $\lambda$'s are all real,  and thereby $a_i$'s are all real. 
   If $a_i \geq 0 ~ \forall ~ i,$ we have  $\rho_{\mathcal{AB}}^{T_{\mathcal{B}}} \geq 0 $.
   Otherwise $ \rho_\mathcal{AB} $ has NPT.
   Now, we 
   can determine whether negative roots of the characteristic equation exist.
    We can write each of these coefficients as a polynomial in $\lambda$s'(eigenvalues of $\rho_{\mathcal{AB}}^{T_{\mathcal{B}}} $) using the Newton-Girard formulae  \cite{Mead}.
    In fact we have : $a_{1} = 1 $  because partial transposition operation is trace preserving; 
    \begin{align}\label{a2} a_{2} = \frac{1}{2} (1 - \sum_{i=1}^{d^2} \lambda_i^2)
   = \text{Tr}(W_2 \rho_{\mathcal{AB}}^{ \otimes 2}),~~ \text{with}, \nonumber\\
  W_2 = \frac{1}{2} \Big[ \mathbb{I}_{d^4 \times d^4} - V^{(2)}  \Big];
  \end{align}

   \begin{align}\label{a3}
   a_3 = \frac{1}{6} \Big(  1 - 3\sum_{i=1}^{d^2} \lambda_i^2 + 2 \sum_{i=1}^{d^2} \lambda_i^3 \Big) \nonumber 
  = \text{Tr}(  W_3 \rho_{\mathcal{AB}}^{\otimes 3}  ),  \nonumber\\
   ~~ \text{with}, ~  W_3 = 
   \frac{1}{6} \Big[~ \mathbb{I}_{d^6 \times d^6} - 3~\mathbb{I}_{d^2 \times d^2}~ \otimes ~ V^{(2)} \nonumber\\
  + ~  \tilde{V}^{(3)}~\otimes~ \tilde{V}^{(3)T} + \tilde{V}^{(3)T}~ \otimes ~ \tilde{V}^{(3)}  \Big] ;
   \end{align}
   
   and,  $ a_4 =  $ \begin{align}
   \frac{1}{24} \bigg( 1 - 6 \sum_{i=1}^{d^2} \lambda_i^2  + 3 \Big( \sum_{i=1}^{d^2} \lambda_i^2  \Big)^2 
   + 8 \sum_{i=1}^{d^2} \lambda_i^3  - 6 \sum_{i=1}^{d^2} \lambda_i^4  \bigg) \nonumber\\  
   =  \text{Tr} \big(  W_4 \rho_{\mathcal{AB}}^{\otimes 4} \big) ~~~~~~~~~~~~~~~~~~~~~~~~~~~~~~~~~~~~~~~~~~~~~ 
   \end{align}
    where $ W_4$ has the same expression as that of  
   $ W^{\text{univ}}$  in Equ. (\ref{univwit}); 
       so on and so forth.
  
   Our argument starts from finding the signature of the coefficients $a_i$ with $i \geq 2$. 
   For most favourable case, $ a_2$ may turn out to be negative and thereby 
   $ \rho_\mathcal{AB} $ has NPT.
   So, two copies of
   the shared state, $\rho_\mathcal{AB}$ is enough to determine NPT-ness while both the inputs 
   $\tau_s $ and $ \omega_t$ are density matrices on Hilbert space  $( \mathbb{C}^d )^{\otimes 2} $.
   But for the general case,  the players have to share `$ i$' number of copies of the state $\rho_{\mathcal{AB }}$ 
   whenever referee wants to know the sign of the coefficient  $a_i$. 
   Also the dimension of the input states ($\tau_s$ and $\omega_t$ to be supplied by the referee) should increase accordingly.
   For the worst case, we have to go upto the coefficient $ a_{d^2}$ for which we need $ d^2 $ copies of $\rho_\mathcal{AB}$
  and $\tau_s $ and $ \omega_t$ will be density matrices on the Hilbert space  $( \mathbb{C}^d )^{\otimes d^2} $.
   As, for {\it any} PPT state $ \rho_\mathcal{AB},$ all of $ \lambda_1, \lambda_2, \lambda_3, ..., \lambda_{d^2} $ are non-negative, all the
  coefficients $ a_2, a_3, ..., a_{d^2} $ are non-negative. So, the operators, $ W_2, W_3, ..., W_{d^2} $ indeed 
  work as witnessing NPT-ness of the state  $ \rho_\mathcal{AB}$.
   
   Note that, these witness operators don't depend on the shared state $\rho_{\mathcal{AB}}$ ,  and so,  we can use the scheme similar to 
that described in section \ref{sectwo} to implement the measurement of these operator in measurement device independent way.
A notable point is that,
for two-qubit case, under partial transposition,  the shared density matrix can have only one negative eigenvalue,
if the state is entangled \cite{augusiak}. Thus for the case of two-qubits, the sign of the coefficient $a_4$ is enough to 
decide the NPT-ness, and thereby entanglement of the state as done in section \ref{sectwo}. We don't need to check the signs of the coefficients $a_2, a_3$.   
But for higher dimensions, that is not the case.  
   
  \subsection{\large Comparison between our MDI scheme and the conventional state tomography :}
  
  In state-tomography, for a two-qudit system, we surely need,  $d^4 -1$ number of measurement settings,  as the 
  number of coefficients in any basis representation ( like generalized Gell-Mann matrices $\Lambda_n$  \cite{12} ) are $ d^4 -1 :$ 
  \begin{align}
   \rho_{d^2 \times d^2} = \frac{1}{d^4} \bigg[ \mathbb{I} \otimes \mathbb{I} + \sum_{n = 1}^{d^4 - 1} r_n \Lambda_n \bigg].
  \end{align} Coefficients $r_n \in \mathbb{R}$, can be obtained from the expectation value of observable $\Lambda_n : $ 
   $ r_n = \text{Tr} \big[~ \rho_{d^2 \times d^2}  \Lambda_n   \big]  $.
   Therefore, for each coefficient, atleast single copy of the state has to be used. 
   In order to reconstruct the 
  state,  we need to know exact values of the coefficients,  their signs only don't give the sufficient information. 
  The erroneous numerical values of these coefficients lead to a wrongly estimated  state, $ \rho_{d^2 \times d^2} $.  
  So, it is not easy to think of a scheme like mesurement device independent tomography for a two-qudit system.

 But in our case, without going into any MDI scheme, to measure $ a_2$ in Equ. (\ref{a2}),  we have to measure the expectation value of 
 a single observable $ V^{(2)} $ only. For $ a_3 $ in Equ. (\ref{a3}), we need to measure, expectation values of two observables
 $ \mathbb{I}_{d^2 \times d^2} \otimes V^{(2)}, ~ \tilde{V}^{(3)} \otimes \tilde{V}^{(3)T} +  \tilde{V}^{(3)T} \otimes \tilde{V}^{(3)}.$
 Similarly for $ a_4, $ we have four observables to measure.  
  So, we need very less number of measurement settings compared to the tomography, at the cost of $n-$copy usage for coefficient $a_n$.
 
 In the universal MDI scheme of finding NPT-ness of any two-qudit state, for the coefficient $ a_n,
 $ we need two general Bell state projectors in Hilbert space $\mathbb{C}^{d^n} \otimes \mathbb{C}^{d^n},$
 and the input density matrices $\tau_s,$ $ \omega_t$ on Hilbert space $ \mathbb{C}^{n^2}.$ These inputs will span the vector space of $W_n$, 
 so, their required number will depend on the number of 
 observables in the expression of $W_n,$ which is in general less than the observable in conventional tomography. 
 For our case,  the sign of the coefficients $a_n$'s are 
  enough to decide whether the shared state $ \rho_\mathcal{AB} $ is entangled or separable.

   \section{\normalsize Witnessing an arbitrary PPT entangled state of a given bipartite system:}
  \label{secfour}
   Existence of PPT entangled state $\rho_{\mathcal{AB }} $ (so called PPT bound entangled states) is well-known whenever 
  $ \text{dim}(\mathcal{H_A}) \times  \text{dim}(\mathcal{H_B})> 6 .$    Evidently our scheme of universal NPT-ness witness, 
  as described in section \ref{secthree}, does not work here to detect entanglement of any such state.  Below we try to argue 
  that a universal entanglement witness operator can not exist to single out entanglement in any PPT state of a 
  given bipartite sytem, through some conjectures.

  \underline{\bf Conjecture I :} 
  If $ \text{dim}(\mathcal{H_A}) \times  \text{dim}(\mathcal{H_B})> 6, $  there can not exist one (or a
finitely many) universal entanglement witness operator detecting entanglement in an arbitrary PPT state $ \rho_\mathcal{AB},$ 
and which can be realized in a MDI way.

Reason behind this conjecture is another conjecture,

\underline { \bf Conjecture II :} There can not exist a universal EW (or, a finitely many EW
operators) for all PPT entangled states of any given bipartite
system

 Let us try to provide some geometrical argument which potentially may give rise to Conjecture II.  
  
  Let $\mathcal{D_{AB}},$ $ \mathcal{S_{AB}},$ $ \mathcal{P_{AB}}$ are respectively be the set of density matrices, 
  separable density matrices, density matrices which are positive semidefinite under partial transposition (PPT), on $\mathcal{H_{AB}}$;
  $\mathcal{\tilde{P}_{AB}}$ be the set of all entangled density matrices (PPTE) in $\mathcal{P_{AB}}$. 
  $ \mathcal{S_{AB}}$ and $ \mathcal{P_{AB}}$ are both convex sets. 
  It is obvious that, $ \mathcal{S_{AB}} \subset \mathcal{P_{AB}} \subset \mathcal{D_{AB}}$ 
  and $ \mathcal{P_{AB}} -  \mathcal{S_{AB}} =  \mathcal{\tilde{P}_{AB}}.$ 
  Note that $\mathcal{\tilde{P}_{AB}}$ is not convex. For $\text{max}\{ dim(\mathcal{H_{A}}),dim(\mathcal{H_{B}}) $\} = 3 together with
  $ \text{min}\{ dim(\mathcal{H_{A}}),dim(\mathcal{H_{B}}) $\} = 2, $ \mathcal{P_{AB}} =  \mathcal{S_{AB}}$. Let $\mathcal{\tilde{P'}_{AB}}$
  is the set of states formed by the convex combination of states $\in ~ \mathcal{\tilde{P}_{AB}}.$
  
  \begin{figure}[center]
 \includegraphics[width=6.5 cm,height= 5.5cm]{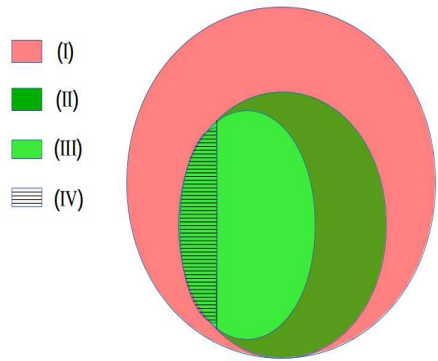}
 \caption{(colour online) Distribution of density matrix spaces on Hilbert space $\mathcal{H_A} \otimes \mathcal{H_B}.$ 
 [ (I) + (II) + (III) + (IV)] $ \equiv \mathcal{D_{AB}},$ the set of all density matrices. 
 (I) $ \equiv \mathcal{D_{AB}} - \mathcal{P_{AB}} $ is the set of all NPT states, $\mathcal{P_{AB}} $ is the set of all states having PPT 
 (II) $ \equiv \mathcal{\tilde{P}_{AB}}$ is the set of all PPT entangled states,
 [ (III) + (IV) ] $ \equiv \mathcal{S_{AB}}$ is the set of all separable states, 
 [ (II) + (III) ] $ \equiv \mathcal{\tilde{P}_{AB}}'$ is the set of convex combination of PPT entangled states. 
 Region (III) clearly depicts the non-empty-ness of $( \mathcal{\tilde{P}_{AB}}'\bigcap \mathcal{S_{AB}}) .$}
\end{figure}
  
  {\bf Edge state :} $ \rho_\mathcal{AB} \in \mathcal{\tilde{P}_{AB}} $ is an edge state iff the state 
  $ (\rho_\mathcal{AB}  + p \sigma_\mathcal{AB}) /( 1 + p ) \in \mathcal{S_{AB}} $
  for all $ p \in (0, 1]$ and all $\sigma_\mathcal{AB} \in  \mathcal{S_{AB}}.$ 
  Edge states lie near the boundary of $ \mathcal{S_{AB}} $ and exist for all 
  $ \text{dim} ( \mathcal{H_A} ,  \mathcal{H_B} ) \geq 3.$

  It is well known that \cite{edge1,edge} there always exist
  atleast two edge states 
  $\rho^{(\alpha)}_\mathcal{AB}, \rho^{(\beta)}_\mathcal{AB} \in \mathcal{\tilde{P}_{AB}}$ such that, their convex combination, 
  $ p \rho^{(\alpha)}_\mathcal{AB} + (1 - p) \rho^{(\beta)}_\mathcal{AB} = \sigma_\mathcal{AB} \in \mathcal{S_{AB}} $ for $ 0 \leq p \leq 1.$
  Hence for, $dim(\mathcal{H_{A}}), dim(\mathcal{H_{B}}) \geq 3, $ $\mathcal{\tilde{P'}_{AB}} \cap  \mathcal{S_{AB}} \neq \phi$ (null set). 
  This is the case for single copy. Fig.2 depicts existence of these edge states in region (II) near the boundary of (II) and (III).  
  
  From Fig.2 it is clear that, there can not exist a hermitian operator
  $ W : \mathbb{C}^d \otimes \mathbb{C}^d \rightarrow  \mathbb{C}^d \otimes \mathbb{C}^d $ such that 
  $ \text{Tr} \big[ W \rho_\mathcal{AB} \big] < 0 $ for all $ \rho_\mathcal{AB} \in  \tilde{\mathcal{P}}_\mathcal{AB} $ while 
    $ \text{Tr} \big[ W \sigma_\mathcal{AB} \big] \geq 0 $ for all $ \sigma_\mathcal{AB} \in \mathcal{S_{AB}}.$
    So, in the single copy case no such universal witness operator $W$ exists to witness entanglement
    in every $\rho_\mathcal{AB} \in \mathcal{\tilde{P}_{AB}}. $

    Now, the question is ---  As in the case of qubits, is it possible to have a ( for finite number of) universal
    witness operator(s) $ W^{\text{univ}}$ which can detect entanglement in any PPT state $ \rho_\mathcal{AB} $ 
    of two-qudits in case $n-$copies of the state are supplied with $ n \geq 2$ ? 
  
 We conjecture that, such a $ W^{\text{univ}}$ does not exist.  
  
  Let's consider, $\mathcal{D}_{\mathcal{A}_n \mathcal{B}_n }$ is the set of all density matrices
  in $ \mathcal{H_A}^{\otimes n} \otimes \mathcal{H_B}^{\otimes n} .$
  $ \mathcal{S'_{AB}}^{(n)}$ is the set formed by the convex combination of the states  $ \sigma_\mathcal{AB}^{ \otimes n},$
  where, $ \sigma_\mathcal{AB} \in \mathcal{S_{AB}} $ and $ \mathcal{\tilde{P'}_{AB}}^{(n)}$ is the set formed by the convex 
  combination of the states  $ \rho_\mathcal{AB}^{ \otimes n},$ where, $ \rho_\mathcal{AB} \in \mathcal{\tilde{P}_{AB}}. $ 
  If $ \mathcal{S}_{\mathcal{A}_n \mathcal{B}_n }$ be the set of all separable states in 
  $ \mathcal{H_A}^{\otimes n} \otimes \mathcal{H_B}^{\otimes n} $, it is evident that,   
   $ \mathcal{S'_{AB}}^{(n)}    \in   \mathcal{S}_{\mathcal{A}_n \mathcal{B}_n } $
   and it is most likely that, $ \mathcal{S}_{\mathcal{A}_n \mathcal{B}_n } \cap \mathcal{\tilde{P'}_{AB}}^{(n)} \neq \phi $ (null set), 
   for all $ n$ with $dim(\mathcal{H_{A}}), dim(\mathcal{H_{B}}) \geq 3$,  
   because of the existence of edge states in  $ \mathcal{H_A}^{\otimes n} \otimes \mathcal{H_B}^{\otimes n} $.

 \underline {\bf Conjecture III:} 
 $ \mathcal{S'_{AB}}^{(n)}  \cap \mathcal{\tilde{P'}_{AB}}^{(n)} \neq \phi $ (null set)
 for all $ n$ with $dim(\mathcal{H_{A}}), dim(\mathcal{H_{B}}) \geq 3.$

 Fig.2 depicts Conjecture III is true for $ n = 1.$
 
   That means, there always exist atleast two states $\rho^{(\alpha)}_\mathcal{AB}, \rho^{(\beta)}_\mathcal{AB} \in \mathcal{\tilde{P}_{AB}}$ such that,
  the convex combination of their $n$ number of copies, 
  \begin{align}\label{pptcon}
  p  \big( \rho^{(\alpha)}_\mathcal{AB} \big)^{\otimes n} + (1 - p) \big( \rho^{(\beta)}_\mathcal{AB} \big)^{\otimes n} 
  = \sum_r   p_r  \big( \sigma^{(r)}_\mathcal{AB} \big)^{\otimes n} 
  \end{align}
  is a density matrix describes a separable state $ \in \mathcal{H_{A}}^{\otimes n } \otimes  \mathcal{H_{B}}^{\otimes n },$
  $ \sigma^{(r)}_\mathcal{AB}  \in \mathcal{S_{AB}},$ and this holds for any $n \in \mathbb{N}.$

  Where $ \mathbb{N}$ is the set of all natural numbers, $ 0 \leq p_r \leq 1 $ forall $ r$ and $ \sum_r p_r = 1$.
  
   \begin{figure}[center]
 \includegraphics[width=7 cm,height= 6 cm]{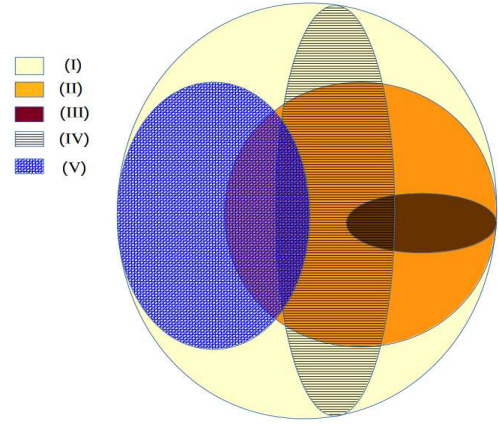}
 \caption{ (colour online) Distribution of density matrix spaces on Hilbert space $\mathcal{H_A}^{\otimes n} \otimes \mathcal{H_B}^{\otimes n}.$ 
  [ (I) $ \bigcup $ (II) $ \bigcup $ (III) $ \bigcup $ (IV) $ \bigcup $ (V) ] $ \equiv \mathcal{D}_{\mathcal{A}_n \mathcal{B}_n }, $ 
  the set of all density matrices.
  (II)$ \equiv \mathcal{S}_{\mathcal{A}_n \mathcal{B}_n }$ is the set of all separable states,
  (III) $ \equiv \mathcal{S'_{AB}}^{(n)}$  is the set of convex combinations of all states like $ \sigma_\mathcal{AB}^{ \otimes n},$ where 
  $ \sigma_\mathcal{AB} \in \mathcal{S_{AB}} $. (IV) $ \equiv \mathcal{\tilde{P'}_{AB}}^{(n)} $ is the convex combination of the states 
  like $ \rho_\mathcal{AB}^{ \otimes n},$ where $ \rho_\mathcal{AB} \in \mathcal{\tilde{P}_{AB}}.$
  (V) contains convex combination of all the states like $ \rho_\mathcal{AB}^{ \otimes n},$ where 
  $ \rho_\mathcal{AB} \in \mathcal{D_{AB}} - \mathcal{\tilde{P}_{AB}},$ set of all NPT states at the single copy level.
  We conjecture about the non-empty-ness of the intersection  (III) $ \bigcap $ (IV), and the empty-ness of, (III) $ \bigcap $ (V) = $ \phi$ (null set). }
\end{figure}

  \vspace{0.8cm}
  
  Validity of conjecture III implies nonexistence of any hermitian operator,
  $$ W_{\mathcal{A}_n \mathcal{B}_n} :\mathcal{H_{A}}^{\otimes n } \otimes  \mathcal{H_{B}}^{\otimes n }
  \rightarrow \mathcal{H_{A}}^{\otimes n } \otimes  \mathcal{H_{B}}^{\otimes n },$$
   for which, $ \text{Tr} \big[ W_{\mathcal{A}_n \mathcal{B}_n}  \big( \rho_\mathcal{AB} \big)^{\otimes n}  \big] < 0 $
  for {\it some} $ \rho_\mathcal{AB} \in \mathcal{\tilde{P}_{AB}} $ 
   together with $ \text{Tr} \big[ W_{\mathcal{A}_n \mathcal{B}_n}  \big( \sigma_\mathcal{AB} \big)^{\otimes n}  \big] \geq 0 $
   for {\it all}  $ \sigma_\mathcal{AB} \in \mathcal{S}_\mathcal{AB}$ --- irrespective of the choice of $n.$

   Thus, validity of conjecture III implies that there can not exist
a universal EW (or, a finitely many EWs) which can detect
entanglement in all the PPT (bound) entangled states of
$ \mathcal{A} + \mathcal{B}$ whenever $d \geq 3.$
   
   This atomatically implies the validity of the conjecture II. 
  It appears to be quite difficult to verify conjecture III directly. One may try to verify whether for some given
  states $ \rho^\alpha_\mathcal{AB} \in \tilde{P}_\mathcal{AB},$ there exist a separable state,
  $ \sigma_\mathcal{AB} \in \mathcal{H_A}^{\otimes n} \otimes \mathcal{H_B}^{\otimes n}$ such that,
  $$ \text{Tr} \Big[  \mathcal{O}~ \sum_\alpha  p_\alpha \big(\rho^{(\alpha)}_\mathcal{AB} \big)^{\otimes n} \Big] = 
   \text{Tr} \Big[  \mathcal{O}~ \sigma_\mathcal{AB}\Big],$$ for a complete set of linearly independent observables
   $ \mathcal{O} :\mathcal{H_A}^{\otimes n} \otimes \mathcal{H_B}^{\otimes n} \rightarrow  \mathcal{H_A}^{\otimes n} \otimes \mathcal{H_B}^{\otimes n}.$
  Even verification of this one may turn out to be difficult. 
  
   This difficulty seems to arise from the well-known fact that the separability problem ( that is to find out whether
   {\it arbitrary} state of a given bipartite quantum system is entangled or separable ) is NP-hard \cite{nphard, nphard2} .

   \section{Noise Analysis for real Experiments : }

      Recently, a few experimental works \cite{10,muha} have been done to implement this MDI protocol 
      to test the entanglement of two particular type of two-qubit states.
      
      We now analyse the effect of possible noises
      in the Bell-state-measurement (BSM) part of their experimental setups.
      Here we introduce three kinds of possible noise (see Fig. 4) : 
      
     {\bf (1) Photon loss in the polarizing beam splitter (PBS) :} We model here the photon loss in a PBS through the action of 
     a white noise - as described below. 
     
     {\bf (2) Error in the angle of rotation in the half wave plate (HWP) :}  HWP 
      rotates the polarization axis of the light vector (which carries the information about the projected state by the measurement operator)
      with respect to it's direction of the propagation of light ray by the angle equals two times of the angle 
      ($\theta$) between the fast axis of HWP and 
      the polarization axis. So, if the angle of rotation be improper, we can it as $2 \theta + \eta $, 
      where $ \eta$ is the error in BSM.
      
     { \bf (3) Detection inefficiency in the PBS and Diode photon-detector parts :  }
           For detection inefficiency, the probability of detection sets reduced by a factor $\xi$ such that
      $0 \leq \xi \leq 1$ and the probability of detection become $\xi P(1,1| \tau_s, \omega_t).$

      For our discussion we have chosen $\ket{\Phi^+}\bra{\Phi^+} $ 
      a Bell state projector as  measurement operator ( as the projector of BSM),
      where $\ket{\Phi^+} = \frac{1}{\sqrt{2}} (\ket{HH} + \ket{VV})$ 
with the consideration that the horizontally polarized state is $\ket{H} = (1 ~~ 0)^T \equiv \ket{0}$ 
and  vertically polarized state is $\ket{V} = (0 ~~ 1)^T \equiv \ket{1}.$

 \begin{figure}[center]\label{pbs}
 \includegraphics[scale = 0.34]{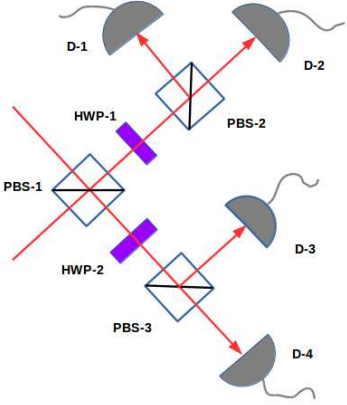}
 \caption{ (colour online) Bell-state projection $ \frac{1}{2} ( \ket{HH} + {VV} )( \bra{HH} + \bra{VV}) $ measurement via coincidence counts.
 Photon propagations are denoted by red-rays. }
\end{figure}

 \begin{figure}[center]\label{hwp}
  \includegraphics[scale = 0.32]{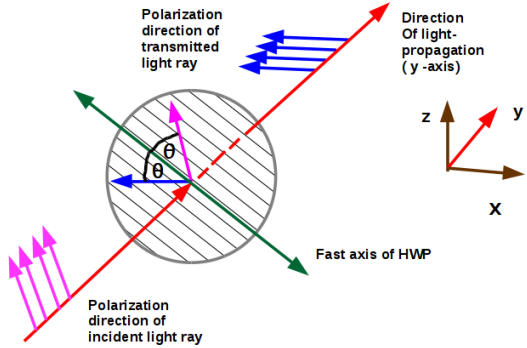}
  \caption{(colour online) Light ray passing through Half wave plate. The polarization axis, rotates by angle $2 \theta$ with respect to the direction of propagation.
  $\theta$ is the angle between the fast axis of the direction of polarization of the incident light ray. The first axis of the HWP lies in the 
  $xz-$plane. }
 \end{figure}

      Under additive white noise (a special case), our actual measurement operator will be,
      $\mu \ket{\Phi^+}\bra{\Phi^+} + \frac{1-\mu}{4} \mathbb{I},$ where $\mu$ is the corresponding visibility.
      
      We assume the direction of propagation of photon is along y axis (denoted by red rays in Fig. 4 and Fig. 5),
      where the reference coordinate frames fixed at every HWP. 
      Thus if the angle of rotation for the polarization axis (shown in Fig. 5) of one player and that coming
      from the referee  are respectively $2 g_1$ and $2 g_2$ then,  {\bf rotated} 
      \begin{align}  \ket{\Phi^+} = e^{-i g_1 \sigma_y} \otimes  e^{-i g_2 \sigma_y} \ket{\Phi^+} ~~~~~~~ \nonumber\\
                               = \cos (g_2 - g_1) \ket{\Phi^+} + \sin (g_2 - g_1) \ket{\Psi^-} ~~~~~~~  
                              \end{align}
 where, $ \ket{\Psi^-} = \frac{1}{\sqrt{2}} ( \ket{ H V} - \ket{V H} ). $ 
 So, if $g_1$ and $g_2$ are the same or $ \pm \pi$ or $\pm 2 \pi $ then there is no error in the HWP. Therefore after passing through noisy PBS and HWP, 
our noisy Bell state projector will be, \begin{align}
                               P =   \mu \cos^2 (g_2 - g_1) \ket{\Phi^+}\bra{\Phi^+} \nonumber\\ 
                                   + \mu  \cos (g_2 - g_1) \sin (g_2 - g_1) \Big(\ket{\Psi^-} \bra{\Phi^+} + \ket{\Phi^+}\bra{\Psi^-} \Big) \nonumber\\
                                   + \mu \sin^2 (g_2 - g_1) \ket{\Psi^-} \bra{\Psi^-} + \frac{1-\mu}{4} \mathbb{I}
                                  \end{align}
Let's take $ g_2 - g_1 = \Delta. $
Then the noisy projection operator shared between $\mathcal{A}$ and a part come from the referee $\mathcal{A_O}$ = 
\begin{align}
 P_\mathcal{A_O A} =   \mu_1 \cos^2 (\Delta_1) \ket{\Phi^+}\bra{\Phi^+} \nonumber\\ 
                                   + \mu_1  \cos (\Delta_1) \sin (\Delta_1) \Big(\ket{\Psi^-} \bra{\Phi^+} + \ket{\Phi^+}\bra{\Psi^-} \Big) \nonumber\\
                                   + \mu_1 \sin^2 (\Delta_1) \ket{\Psi^-} \bra{\Psi^-} + \frac{1-\mu_1}{4} \mathbb{I}
                                  \end{align}
Similarly for $\mathcal{B}$ and $\mathcal{B_O}$, 
\begin{align}
 P_\mathcal{B  B_O} =   \mu_2 \cos^2 (\Delta_2) \ket{\Phi^+}\bra{\Phi^+} \nonumber\\ 
                                   + \mu_2  \cos (\Delta_2) \sin (\Delta_2) \Big(\ket{\Psi^-} \bra{\Phi^+} + \ket{\Phi^+}\bra{\Psi^-} \Big) \nonumber\\
                                   + \mu_2 \sin^2 (\Delta_2) \ket{\Psi^-} \bra{\Psi^-} + \frac{1-\mu_2}{4} \mathbb{I}
                                  \end{align}
                                  
 Here, $ \mu_1$ and $\mu_2 $ are the visibilities of the PBS's of the party $ \mathcal{A}-$ side and $ \mathcal{B}-$side
 respectively in the noisy measurement of $ \ket{\Phi^+}.$ $ \Delta_1$ and $ \Delta_2 $ are the corresponding errors 
 in rotational angle in HWP in $ \mathcal{A}-$ side and $ \mathcal{B}-$side respectively.

          So, our desired (modified) quantity will be,  \begin{align}\label{modifiedi}
                                           I_\text{mod}(  \rho_\mathcal{AB})  = ~~~~~~~~~~~~~~~~~~~~~~~~~~  \nonumber\\
                                           \sum_{s,t} \beta_{s,t} ~\xi~~
                                             \text{Tr} \Big[ (P_\mathcal{A_O A} \otimes P_\mathcal{B B_O }) 
                                             (\tau_s \otimes \rho_\mathcal{AB} \otimes \omega_t)\Big]~~~~~~ \nonumber\\
                                             =   \xi \bigg[ \mu_1 \mu_2 \cos^2 ( \Delta_1) \cos^2 ( \Delta_2 )
                 + \frac{\mu_1( 1 - \mu_2)}{4} \cos^2 ( \Delta_1)  \nonumber\\
    + \frac{\mu_2( 1 - \mu_1)}{4} \cos^2 ( \Delta_2)  
    + \frac{(1 -\mu_1)( 1 - \mu_2)}{16}\bigg]  \nonumber\\
    ~ I (  \rho_\mathcal{AB}) + \text{additional term}. 
                                            \end{align}
The additional term is in Equ. (\ref{app3}) and (\ref{app4}) of appendix \ref{appen3}. 
         
          Now, for simplicity, let us assume that $\mu_1 = \mu_2 = \mu$ and $ \Delta_1 = \Delta_2 = \Delta.$
          
           In the case of a two-qubit shared state $\rho_\mathcal{AB},$ the input states $ \tau_s$ and $ \omega_t$ takes
           the forms,
            \begin{widetext}
            \begin{align}\label{statechoice}
                                                              \tau_s = \frac{\mathbb{I} + \vec{r}_s.\vec{\sigma}}{2}, ~~ \omega_t =    \frac{\mathbb{I} + \vec{r}_t.\vec{\sigma}}{2},~~
                                                              \rho_\mathcal{AB} = \frac{1}{4} \Big( \mathbb{I} \otimes \mathbb{I} + \sum_{i=1}^3 a_i~~\mathbb{I} \otimes \sigma_i 
                                                              + \sum_{i=1}^3 b_i~~\sigma_i \otimes \mathbb{I} + \sum_{i,j = 1}^3 c_{ij}~~ \sigma_i \otimes \sigma_j \Big) \\
                                                              ~~~\text{with}, ~~~  \vec{r}_{s,t} = (x_{s,t} ,~ y_{s,t},~ z_{s,t}) \in \mathbb{R}^3,~
                                                              |\vec{r}_{s,t}| \leq 1,~ ( a_1, a_2, a_3 ) \in \mathbb{R}^3,~ ( b_1, b_2, b_3) \in \mathbb{R}^3 ~ \nonumber\\
                                                              \text{and}~ ( c_{ij})_{i,j =1}^3 ~ \text{is a real}~ 3 \times 3 ~ \text{matrix.}\nonumber
                                                              \end{align} \end{widetext}

                                                         The corresponding form of 
                                                         $ I_\text{mod}(\rho_\mathcal{AB})$ is given in, appendix [\ref{appen3}]

 According to ref. \cite{muha}, the input states are given in table -I,

 \begin{tabular}{ |p{ 1.7 cm}||p{1.7 cm}|p{1.7 cm}|p{1.7 cm}|  }
 \hline
 Input states $ \tau_s, \omega_t $ & Bloch vectors $ \vec{r}_s, \vec{r}_t$ \\
 \hline
$ \ket{H}\bra{H} $ & (~0,~0,~1) \\
 $\ket{V}\bra{V} $  &  (~0,~0,-1) \\
 $\ket{D} \bra{D} $  & (~1,~0,~0)  \\
 $\ket{\bar{D}} \bra{\bar{D}} $  & (-1,~0,~0) \\
 $\ket{L} \bra{L} $ & (~0,~1,~0) \\
 $\ket{R} \bra{R}$ & (~0,-1,~0) \\
  \hline
\end{tabular} \\\\
$~~~~ ${\bf Table - I : }
Input states used in the ref. \cite{muha}
for Alice and Bob's side and their corresponding bloch vectors.\\

The corresponding payoffs are :
\begin{align}\label{betaval}
 \beta_{H,H} = \beta_{V,V} =  \beta_{D,D} = \beta_{\bar{D},\bar{D}} = \beta_{L,L} = \beta_{R,R} = \frac{1}{3},\nonumber\\
 \beta_{H,V} = \beta_{V,H}  = \beta_{D,\bar{D}} = \beta_{\bar{D},D} = \beta_{L,R} = \beta_{R,L}  = - \frac{1}{6}. 
\end{align}
Other $\beta_{s,t}$ are equal to zero.

The shared state is, $\rho_\mathcal{AB} =  p \ket{\Psi^-} \bra{\Psi^-} + \frac{1-p}{4}~\mathbb{I}$
which is separable iff $ 0 \leq p \leq \frac{1}{3}.$ Thus in Equ. (\ref{statechoice}),
$c_{11} = c_{22} = c_{33} = -p.$ $c_{kl} = 0; $ when $ k \neq l;$  $a_k = b_k =0 $   $ \forall k,l \in \{ 1,2,3\}.$ 

Using Equ. (\ref{betaval}) and table-I we have, 
\begin{align}
      \sum_{s,t} \beta_{s,t} = 1,~~~~\nonumber\\ 
    \sum_{s,t} \beta_{s,t}~ x_s x_t =    \sum_{s,t} \beta_{s,t}~ y_s y_t = 
      \sum_{s,t} \beta_{s,t}~ z_s z_t = 1, \nonumber\\
        \sum_{s,t} \beta_{s,t} x_s y_t =  
        \sum_{s,t} \beta_{s,t} x_s z_t =  \sum_{s,t} \beta_{s,t} y_s x_t  \nonumber\\
         = \sum_{s,t} \beta_{s,t} y_s z_t =  \sum_{s,t} \beta_{s,t} z_s x_t =     \sum_{s,t} \beta_{s,t}  z_s y_t = 0.      
                \end{align}

       Therefore, according to Equ. (\ref{modifiedi})  ~ ( see Equ. (\ref{modwit}) in  Appendix \ref{appen3} ),
                 \begin{align}  I_\text{mod} (\rho_\mathcal{AB}) = \frac{\xi}{4} \big[  1 -  p \mu^2 - 2 p \mu^2 \cos (4 \Delta) \big] \end{align} 
                 For the allowed values of the noise parameters $ \xi, \mu , \Delta $ we have,  $ 0 \leq  I_\text{mod}(\rho_\mathcal{AB}) \leq \frac{1}{3} .$
                     Hence for these kinds of errors no separable state will never be detected as entangled state.

    Ref. \cite{10} deals with the entangled state, $\rho = (1-r) \ket{\Psi^-} \bra{\Psi^-} 
    + \frac{r}{2} ( \ket{HH}\bra{HH} + \ket{VV}\bra{VV}) $ with $ 0 \leq r \leq 1.$ 
    $ \rho_\mathcal{AB} $ is separable for $ \frac{1}{2} \leq r \leq 1.$
    Here, for this case, $c_{11} = c_{22} = (r-1) $ and $c_{33} = 2 r - 1. $ $c_{kl} = 0; $
    when $ k \neq l;$  $a_k = b_k = 0,$  $ \forall k,l \in \{ 1,2,3\}.$
       
        Here we are going to use the data of the Table II of the  Supplementary section of ref. \cite{10}.
  
  for that case, 

  \begin{align}
             \sum_{s,t} \beta_{s,t}   = 1, ~~~~~~~~~~~~~~~~~~~~~~~~~ \nonumber\\
            \sum_{s,t} \beta_{s,t} x_s x_t = 
             \sum_{s,t} \beta_{s,t} y_s y_t  = 1,     
                   \sum_{s,t} \beta_{s,t} z_s z_t  = 1, \nonumber\\ 
                  \sum_{s,t} \beta_{s,t} z_s x_t =  \sum_{s,t} \beta_{s,t}  z_s y_t =              
                  \sum_{s,t} \beta_{s,t} x_s y_t = \nonumber\\  \sum_{s,t} \beta_{s,t} x_s z_t =  
                  \sum_{s,t} \beta_{s,t} y_s x_t =   \sum_{s,t} \beta_{s,t} y_s z_t = 0, \nonumber\\
              \end{align}
  Therefore,  according to Equ. (\ref{modifiedi}),~ ~~ ( see Equ. (\ref{modwit}) in Appendix \ref{appen3} ),
              \begin{align}               I_\text{mod}(\rho_\mathcal{AB}) = \frac{\xi}{4} [(3 r - 2) \mu^2 \cos(4 \Delta) +  \mu^2 (r -1) + 1 ]. \end{align}
                For $\frac{1}{2} \leq r \leq 1 $ and for allowed values of noise parameters $\xi, \mu, \Delta$, we have
                 $0 \leq  I_\text{mod}(\rho_\mathcal{AB}) \leq \frac{1}{2}. $  So, as in the case of Ref. \cite{muha}, separable states will never detected
                  as entangled.

      \section{Conclusion:} \label{secsix} 
     Using the prescription of Auguisiak et al. \cite{augusiak} for witnessing entanglement in a unknown state of two-qubits, 
     we have extended here the measurement device independent entanglement witness scheme of Branciard et al. \cite{Gisin-mdiew}
    to a universal measurement device independent entanglement witness scheme for two-qubit states, the caveat being 
    that we need four copies of the state at a time and the referee should supply the input statesfrom sixteen dimension
    Hilbert spaces of Alice and Bob. In an aim to extend this result to higher dimension, we provided here a measurement
    device independent scheme for universally witnessing NPT-ness of a unknown state of any given bipartite system - provided 
    many copies of the state is being supplied. Of course, that number of copies is much less than that required for 
    state tomography.    
    We conjectured that, there doesn't exist a single or a finite number of universal witness operators for witnessing 
    entanglement in an arbitrary PPT state of a given bipartite system. Our noise analysis of the Bell state measurement
    scenario in both the experimental demonstration of MDIEW \cite{10, muha} are in conformity with the MDI of EW of 
    Branciard et al. \cite{Gisin-mdiew}.
    \vspace{0.3cm}
   
   ACKNOWLEDGEMENT: The authors would like to thank Sandeep K. Goyal and Manik Banik for useful discussion on the present work.

\begin{widetext}   
   
  ~~~~~~~~~~~~~~~~~~~~~~~~~~~~~~~~~~~~~~~~ ~~~~~~~~~~~~~~~~~~~~~~~ {\large{\bf APPENDIX}}

   \subsection{ \normalsize Inputs and measurement operators are noisy}\label{secfive}
    An usual apparatus is generally influenced by ambient noise. In \cite{Gisin-mdiew},  the players and referee trust the preparation of the input states. 
    So, the noise in $\tau_s$ and 
   $\omega_t$ (if any), is supposed to be known by the referee (or, even by the players). Some noise may affect the actual shared state
   $ \rho_\mathcal{AB},$ but as we are going to detect it, we will consider the noise induced shared state as the actual state $ \rho_\mathcal{AB},$
   need to be detected.

   Any general $d-$dimensional noise can be expressed as $\frac{1}{d}\vec{n}.\vec{\Lambda}_{d\times d} $ where $n^i$ and 
   $\Lambda^i$ are the $i^{th}$ components of generalized Bloch vector and Gell-Mann matrix in $d$-dimension respectively,
   $\Lambda^0 =\mathbb{I}_{d \times d} $ \cite{12}. If the visibility $\mu $ of the actual input state is
   same for all input states, and assume the visibility remains constant throughout the experiment, 
   then the players (Alice and Bob) would receive the input states, 
   
  \begin{align}
    \tau_s' = \mu \tau_s + \frac{(1 - \mu)}{d}\vec{n}_1.\vec{\Lambda}_{d\times d} 
    = \Big[\frac{(1-\mu)}{d}\vec{n}_1 + \frac{\mu}{d}\text{Tr}(\tau_s \vec{\Lambda}_{d\times d} ) \Big].\vec{\Lambda}_{d \times d},\nonumber
~ \text{and}~~~  \omega_t' =  \Big[\frac{(1-\mu)}{d}\vec{n}_2 + \frac{\mu}{d}\text{Tr}(\omega_t \vec{\Lambda}_{d\times d})\Big]
    .\vec{\Lambda}_{d \times d} \nonumber \end{align}  instead of $\tau_s $ and $\omega_t $ respectively . 
If the referee does not aware about the noise mixed with the inputs, there will be mismatch of the value of $\mathcal{I} ( \rho_{\mathcal{AB}}) $
     from it's actual value if the refree knows the noise.

     In the game, the form of $W$ is known to the referee, he calculates the pay-off functions, $\beta_{s,t}$ according to the inputs [\ref{beta}].   
     If referee uses $\tau_s$  and $\omega_t$ as the inputs instead of $\tau_s'$  and $\omega_t'$, his calculated 
    $\beta_{s,t}$ will be different than the actual that will give the actual value of $ I(\rho_{\mathcal{AB}}) $.
   
    Modified witness operator looks like,
    \begin{align}
    W'= \sum_{s,t} \beta_{s,t}  \tau_s^{'T}\otimes\omega_t^{'T} = \frac{\mu^2}{d^2} \sum_{s,t} \beta_{s,t}  \tau_s^{T}\otimes\omega_t^{T} 
    + ~~ \text{additional noise terms.} 
   \end{align}
   So, by knowing the amount of noise induced in inputs, referee can calculate the modified value of $  \beta_{s,t}$, which comes from the  
   linear equation generated by, $ \text{Tr} [W'  ( \tau_s^{'T}\otimes\omega_t^{'T} ) ],$ for all $ s, t.$ 
   
    So, we can claim that if the referee knows about the character of the noise in inputs,  entanglement determination shouldn't be 
    erroneous.

    \vspace{0.5cm}
    In general, the measurement operators can also noisy, and the character of the noise 
    is not supposed to be known by the referee nor by the players. But that noise can obviously affect the 
    final decision based on the value of $ I(\rho_\mathcal{AB})$.   
    If a general additive noise,
    $\frac{1}{d^2}$ $\vec{m}.\vec{\Lambda}_{d^2 \times d^2}$
    is added with the maximally entangled 
    projector $ \frac{1}{d}\sum_{i,j=1}^d \ket{ii}\bra{jj}, $  then with the visibility $\nu$ of the original measurement operator,
    the actual measurement operator appears as 
    \begin{align}
    \frac{\nu}{d}\sum_{i,j} \ket{ii}\bra{jj} +
    \frac{(1 - \nu)}{d^2}\vec{m}.\vec{\Lambda}_{d^2 \times d^2},\nonumber
    \end{align}
    provided this is a positive operator,  that is $ \forall $ $\ket{\chi}$ we have,  
    
    \begin{align}\bra{\chi} \bigg( \frac{\nu}{d}\sum_{i,j} \ket{ii}\bra{jj} +
    \frac{(1 - \nu)}{d^2}\vec{m}.\vec{\Lambda}_{d^2 \times d^2} \bigg) 
    \ket{\chi} \geq 0,\nonumber \end{align}
    
    In the standard basis,$ \{\ket{i,j} \}$, we can write 
    \begin{align} \vec{\Lambda}_{d^2 \times d^2} = \sum_{a,b,p,q} \bra{a,b}\vec{\Lambda}\ket{p,q}  \ket{a,b}\bra{p,q} \nonumber \end{align}

   For notational convenience,
    in place of $\rho_{\mathcal{AB}}$,  we will be using $\rho $ here
    and also $\Lambda $ for  $ \Lambda_{d^2 \times d^2}$ .

   \begin{align}\label{noisemea}
 P_{\rho}(1,1|s,t) = \text{Tr}\Bigg[ \Bigg\{ \bigg(\frac{\nu}{d}\sum_{i,j} \ket{ii}\bra{jj} + \nonumber
     \frac{(1 - \nu)}{d^2}\vec{m_1}.\vec{\Lambda}\bigg)\otimes \nonumber %\\
    \bigg(\frac{\nu}{d}\sum_{u,v} \ket{uu}\bra{vv}+ \frac{(1 - \nu)}{d^2}\vec{m_2}.\vec{\Lambda}\bigg)\Bigg\} \nonumber
    \bigg( \tau_s \otimes\rho\otimes \omega_t\bigg)\Bigg] \nonumber\\
   =\frac{\nu^2}{d^2}\text{Tr}\bigg[ \Big( \tau_s^T\otimes\omega_t^T \Big) ~ \rho \bigg] 
   + \frac{\nu(1-\nu)}{d^3}\text{Tr}\bigg[ \Big( \tau_s^T\otimes \text{Tr}_{\mathcal{B_O}}[ \nonumber
 (\vec{m_2}.\vec{\Lambda})(\mathbb{I}\otimes\omega_t)] \Big) ~ \rho\bigg] ~~~~~~~~~~~~~~~~~~~~~~~~~~~~~~  \nonumber \\
     +\frac{\nu(1-\nu)}{d^3}\text{Tr}\bigg[  \Big( \text{Tr}_{\mathcal{A_O}}[(\tau_s\otimes\mathbb{I})\nonumber
 (\vec{m_1}.\overrightarrow{\Lambda})]\otimes\omega_t^T \Big) ~ \rho \bigg] 
   +\frac{(1-\nu)^2}{d^4}\text{Tr}\bigg[  \Big( \text{Tr}_{\mathcal{A_O}}[(\tau_s\otimes\mathbb{I})\vec{m_1}.\vec{\Lambda}] \otimes \nonumber
   \text{Tr}_{\mathcal{B_O}}[\vec{m_2}.\vec{\Lambda}(\mathbb{I}\otimes\omega_t)] \Big)~ \rho\bigg] \nonumber \\
   = \frac{1}{d^2}\text{Tr}\bigg[\tau_s^{''T}\otimes\omega_t^{'' T}\rho\bigg].~~~~~~~~~ ~~~~~~~~~~~~~~~~~~~~~~~~~~~~~~~~~~~~~~~~~~~~~~~~~ 
   \end{align}
  
   If in (\ref{noisemea}) $\vec{m}_1.\vec{\Lambda}$ = $\vec{m}_2.\vec{\Lambda} $ =
     $ \mathbb{I}_{d \otimes d} \otimes \mathbb{I}_{d \otimes d}$ ,  will be denoted here as $\mathbb{I}\otimes \mathbb{I}$,
    and if define $\vec{\Lambda}$ = $(\mathbb{I},\vec{\Gamma}) $, then

      \begin{align} \tau_s^{''T}\otimes\omega_t^{''T} ~~~~~~~  =  ~~~~~~~~~  
   \frac{\nu^2}{d^2}\tau_s^T\otimes\omega_t^T +  \frac{\nu(1-\nu)}{d^3}\tau_s^T\otimes tr_{\mathcal{B_O}}\bigg[
     (\mathbb{I}\otimes \mathbb{I})(\mathbb{I}\otimes\omega_t)\bigg] \nonumber\\
     +\frac{\nu(1-\nu)}{d^3}tr_{\mathcal{A_O}}\bigg[(\tau_s\otimes\mathbb{I})
     (\mathbb{I}\otimes \mathbb{I}) \bigg]\otimes\omega_t^T
     +\frac{(1-\nu)^2}{d^4} tr_{\mathcal{A_O}}\bigg[(\tau_s\otimes\mathbb{I})\mathbb{I}\otimes \mathbb{I} \bigg]\otimes
      \text{Tr}_{\mathcal{B_O}} \bigg[\mathbb{I}\otimes \mathbb{I}(\mathbb{I}\otimes\omega_t)\bigg] \nonumber\\
    = \frac{\nu^2}{d^2}\tau_s^T\otimes\omega_t^T +  
 \frac{\nu(1-\nu)}{d^3}\tau_s^T\otimes\mathbb{I}
      +\frac{\nu(1-\nu)}{d^3}\mathbb{I}\otimes\omega_t^T
     +\frac{(1-\nu)^2}{d^4}\mathbb{I}\otimes\mathbb{I}\nonumber\\
    =\frac{1}{d^4}\mathbb{I}\otimes\mathbb{I}
     +\frac{\nu}{d^4}\Big(\mathbb{I}\otimes\vec{\Gamma}\Big).tr(\vec{\Gamma}\omega_t^T)
     +\frac{\nu}{d^4}\Big(\vec{\Gamma}\otimes\mathbb{I}\Big).tr(\vec{\Gamma}\tau_s^T)
     +\frac{\nu^2}{d^4}\Big(\vec{\Gamma}.tr(\vec{\Gamma}
     \tau_s^T)\Big)
     \otimes \Big( \vec{\Gamma}.tr(\vec{\Gamma} \omega_t^T)\Big)\nonumber\\
    = \frac{1}{d^4} \big(\mathbb{I} + \nu \vec{\Gamma}.tr(\vec{\Gamma} \tau_s^T)\big)
   \big(\mathbb{I} + \nu \vec{\Gamma}.tr(\vec{\Gamma} \omega_t^T)\big) ~~~~~~~~~~~~~~~~~~~~~~~~~~~~~~~~~ 
  \end{align}
   
    Thus this is just a shrinking operation of the generalized Bloch vector $\vec{\Gamma}$.
   
    If the referee is unaware about the noise then, there is deviation from the actual results.So,
    referee will use the $\beta $ functions same as that obtained without noise.In presence of noise in 
    inputs and measurement operator,the witness operator form becomes,
    \begin{align}
     \sum_{s,t} \beta_{s,t} \Bigg[\frac{\nu^2}{d^2}
    \Big(\mu \tau_s + \frac{(1 - \mu)}{d}\vec{n_1}.\vec{\Lambda}_{d\times d}\Big)^T
    \otimes
    \Big( \mu \omega_t + \frac{(1 - \mu)}{d}\vec{n_2}.\vec{\Lambda}_{d\times d}\Big)^T  ~~~~~~~~~~~~~~~~~~~~~~ \nonumber\\ 
     + \frac{\nu(1-\nu)}{d^3}
     \Big( \mu \tau_s + \frac{(1 - \mu)}{d}\vec{n}_1.\vec{\Lambda}_{d\times d}\Big)^T\otimes tr_{\mathcal{B_O}}\Big[
     (\vec{m}_2.\vec{\Lambda})(\mathbb{I}\otimes
     \Big( \mu\omega_t + \frac{(1 - \mu)}{d}\vec{n}_2.\vec{\Lambda}_{d\times d}\Big))\Big]     ~~~~~~~~~~~~~~ \nonumber\\
     +\frac{\nu(1-\nu)}{d^3}
     tr_{\mathcal{A_O}}\Big[(\Big(\mu \tau_s + \frac{(1 - \mu)}{d}\vec{n}_1.\vec{\Lambda}_{d\times d}\Big)\otimes\mathbb{I})
     (\vec{m}_1.\vec{\Lambda})\Big]\otimes
     \Big(\mu \omega_t + \frac{(1 - \mu)}{d}\vec{n}_2.\vec{\Lambda}_{d\times d}\Big)^T
 ~~~~~~~~~~~~~~  \nonumber\\
     +\frac{(1-\nu)^2}{d^4}\bigg\{tr_{\mathcal{A_O}}\Big[(\Big(\mu \tau_s + \frac{(1 - \mu)}{d}\vec{n}_1.\vec{\Lambda}_{d\times d}
 \Big)\otimes\mathbb{I})\vec{m}_1.\vec{\Lambda}\Big]\otimes tr_{\mathcal{B_O}}\Big[\vec{m}_2.\vec{\Lambda}(\mathbb{I}\otimes
       \Big( \mu \omega_t + \frac{(1 - \mu)}{d}\vec{n}_2.\vec{\Lambda}_{d\times d}\Big))\Big]\bigg\}\Bigg]
    \equiv W''  \text{(say)} \end{align} .

  Therefore, to get the correct result, in the expression of $W$, given by the equation [\ref{beta}], 
  instead of $\tau_s^T \otimes \omega_t^T$ referee should use 
   \begin{align} \tau_s^{'' T}\otimes\omega_t^{'' T} =  
   \frac{\nu^2}{d^2}\tau_s^T\otimes\omega_t^T +
    \frac{\nu(1-\nu)}{d^3}\tau_s^T\otimes \text{Tr}_{\mathcal{B_O}}\bigg[
    (\vec{m_2}.\vec{\Lambda})(\mathbb{I}\otimes\omega_t)\bigg]  \nonumber\\
    +\frac{\nu(1-\nu)}{d^3}\text{Tr}_{\mathcal{A_O}}\bigg[(\tau_s\otimes\mathbb{I})
    (\vec{m_1}.\vec{\Lambda}) \bigg]\otimes\omega_t^T
    +\frac{(1-\nu)^2}{d^4} \text{Tr}_{\mathcal{A_O}}\bigg[(\tau_s\otimes\mathbb{I})\vec{m_1}.\vec{\Lambda} \bigg]\otimes
     \text{Tr}_{\mathcal{B_O}} \bigg[\vec{m_2}.\vec{\Lambda}(\mathbb{I}\otimes\omega_t)\bigg] \end{align}

   By knowing the character of the noise mixed with the measurement operator, the referee can change 
   the $\beta_{s,t}$ values and in this way $ I(\rho)$ remains proportional to $\text{Tr}(W \rho$).
   If  he don't know that, then the error will be proportional to $\text{Tr} [  ( W'' - W ) \rho],$ where, 
   \begin{align}
    W'' - W = \sum_{s,t} \beta_{s,t} \Big[ \tau_s^{''T}\otimes\omega_t^{''T} - \tau_s^{T}\otimes\omega_t^{T} \Big]
   \end{align}
and, $\beta_{s,t}$ are calculated according to the equation [\ref{beta}].

   With noisy input states and measurement operator
  \begin{align}\label{measure}
  I(\rho) = I(\nu , \mu ,\vec{m}_1,\vec{m}_2,\vec{n}_1,\vec{n}_2,\rho)\nonumber 
 & = I(1,1,\vec{0},\vec{0},\vec{0},\vec{0},\rho) \nonumber \\
  + \text{Tr}\Big[\rho \big(W''- \sum_{s,t,a,b} \beta_{s,t}^{a,b} \tau_s^T \otimes\omega_t^T \big)\Big]
  \end{align}

 So noise (both in preparation as well as measurement) can, in principle, degrade the quality of measurement device independent
      implementation of the entanglement witness operator.

  \subsection{ \normalsize Expression for $I_\text{mod}(  \rho_\mathcal{AB}) $ } \label{appen3}
 
 Our original witnessing function was,  \begin{align}\label{app1}
    I (  \rho_\mathcal{AB})  = \sum_{s,t} \beta_{s,t}
                                             \text{Tr} \Big[ \Big(\ket{\Phi^+}\bra{\Phi^+} \otimes \ket{\Phi^+}\bra{\Phi^+} \Big) 
                                             (\tau_s \otimes \rho_\mathcal{AB} \otimes \omega_t)\Big]
  \end{align}
 and the modified form of that is,
 \begin{align}\label{app2}
    I_\text{mod}(  \rho_\mathcal{AB})  = \sum_{s,t} \beta_{s,t} ~\xi~~
                                             \text{Tr} \Big[ (P_\mathcal{A_O A} \otimes P_\mathcal{B B_O }) 
                                             (\tau_s \otimes \rho_\mathcal{AB} \otimes \omega_t)\Big]
  ,\end{align} where the modified measurement operators are given by,
  \begin{align}\label{app3}
    P_\mathcal{A_O A} \otimes  P_\mathcal{B  B_O} = \mu_1 \mu_2 \cos^2 (\Delta_1) \cos^2 (\Delta_2) \ket{\Phi^+}\bra{\Phi^+} \otimes \ket{\Phi^+}\bra{\Phi^+} \nonumber\\
  + \mu_1 \mu_2 \cos^2 (\Delta_1) \cos(\Delta_2) \sin(\Delta_2) \ket{\Phi^+}\bra{\Phi^+} \otimes \Big(\ket{\Psi^-} \bra{\Phi^+} + \ket{\Phi^+}\bra{\Psi^-} \Big) \nonumber\\
  + \mu_1 \mu_2 \cos^2 (\Delta_1) \sin^2 (\Delta_2) \ket{\Phi^+}\bra{\Phi^+} \otimes \ket{\Psi^-}\bra{\Psi^-} \nonumber\\
  + \frac{\mu_1 (1- \mu_2)}{4} \cos^2 (\Delta_1) \ket{\Phi^+}\bra{\Phi^+} \otimes \mathbb{I} \nonumber\\
  + \mu_1 \mu_2 \cos(\Delta_1) \sin(\Delta_1) \cos^2 (\Delta_2) \Big(\ket{\Psi^-} \bra{\Phi^+} + \ket{\Phi^+}\bra{\Psi^-} \Big) \otimes \ket{\Phi^+}\bra{\Phi^+} \nonumber\\
  + \mu_1 \mu_2 \cos(\Delta_1) \sin(\Delta_1) \cos(\Delta_2) \sin(\Delta_2)  \nonumber\\
  \Big(\ket{\Psi^-} \bra{\Phi^+} + \ket{\Phi^+}\bra{\Psi^-} \Big) \otimes  \Big(\ket{\Psi^-} \bra{\Phi^+} + \ket{\Phi^+}\bra{\Psi^-} \Big)\nonumber\\
  + \mu_1 \mu_2 \cos(\Delta_1) \sin(\Delta_1) \sin^2 (\Delta_2) \Big(\ket{\Psi^-} \bra{\Phi^+} + \ket{\Phi^+}\bra{\Psi^-} \Big) \otimes \ket{\Psi^-}\bra{\Psi^-} \nonumber\\
  + \frac{\mu_1 (1 - \mu_2)}{4} \cos(\Delta_1) \sin(\Delta_1)  \Big(\ket{\Psi^-} \bra{\Phi^+} + \ket{\Phi^+}\bra{\Psi^-} \Big) \otimes\mathbb{I} \nonumber\\
  + \mu_1 \mu_2 \sin^2(\Delta_1) \cos^2 (\Delta_2) \ket{\Psi^-}\bra{\Psi^-} \otimes \ket{\Phi^+} \bra{\Phi^+} \nonumber\\
  + \mu_1 \mu_2 \sin^2(\Delta_1) \cos(\Delta_2) \sin(\Delta_2) \ket{\Psi^-}\bra{\Psi^-} \otimes \Big(\ket{\Psi^-} \bra{\Phi^+} + \ket{\Phi^+}\bra{\Psi^-} \Big) \nonumber\\
  +  \mu_1 \mu_2 \sin^2(\Delta_1) \sin^2(\Delta_2)  \ket{\Psi^-}\bra{\Psi^-} \otimes  \ket{\Psi^-}\bra{\Psi^-} \nonumber\\
  + \frac{\mu_1 (1-\mu_2)}{4} \sin^2 (\Delta_1)  \ket{\Psi^-}\bra{\Psi^-} \otimes \mathbb{I}  
  + \frac{(1-\mu_1) \mu_2}{4} \cos^2 (\Delta_2) \mathbb{I} \otimes \ket{\Phi^+}\bra{\Phi^+} \nonumber\\
  + \frac{(1-\mu_1) \mu_2}{4} \cos(\Delta_2)\sin(\Delta_2) \mathbb{I} \otimes \Big(\ket{\Psi^-} \bra{\Phi^+} + \ket{\Phi^+}\bra{\Psi^-} \Big) \nonumber\\
 + \frac{(1-\mu_1) \mu_2}{4} \sin^2 (\Delta_2) \mathbb{I} \otimes \ket{\Psi^-}\bra{\Psi^-} +  \frac{(1-\mu_1)(1-\mu_2)}{16} \mathbb{I} \otimes \mathbb{I} 
 \end{align}
  
  Therefore, from Equ. (\ref{app1}), (\ref{app2}) and (\ref{app3}) we have,
  \begin{align}\label{app4}
    I_\text{mod}(  \rho_\mathcal{AB})  =   \xi \bigg[ \mu_1 \mu_2 \cos^2 ( \Delta_1) \cos^2 ( \Delta_2 ) + \frac{\mu_1( 1 - \mu_2)}{4} \cos^2 ( \Delta_1) \nonumber\\
    + \frac{\mu_2( 1 - \mu_1)}{4} \cos^2 ( \Delta_2)  + \frac{(1 -\mu_1)( 1 - \mu_2)}{16}\bigg] 
    ~ I (  \rho_\mathcal{AB}) + \text{additional term} 
    \end{align}
  Thus, both multiplicative and additive noises are present in the system.
  
  Now, using Equ. (\ref{statechoice}) and (\ref{app4}), we have
  [ for the choices $ \mu_1 = \mu_2 = \mu $ and $ \Delta_1 = \Delta_2 = \Delta$ ],
   \begin{align} \label{modwit}
                                                                I_\text{mod}(  \rho_\mathcal{AB}) = ~\frac{\xi}{8} \sum_{s,t} \beta_{s,t} \Big[  2 a_3 \mu  \sin (2 \Delta ) x_t  
                                                                  + 2 a_1 \mu  \cos (2 \Delta ) x_t-2 a_2 \mu  y_t  -2 a_1 \mu  \sin (2 \Delta ) z_t \nonumber\\
                                                                  +2 a_3 \mu  \cos (2 \Delta ) z_t+2 b_1 \mu (\cos (2 \Delta ) x_s+\sin (2 \Delta ) z_s) -2 b_3 \mu  (\sin (2 \Delta ) x_s-\cos (2 \Delta ) z_s)\nonumber\\
                                                                 -2 b_2 \mu  y_s-2 c_{33} \mu ^2 \sin ^2(2 \Delta ) x_s x_t +c_{13} \mu ^2 \sin (4 \Delta ) x_s x_t -c_{31} \mu ^2 \sin (4 \Delta ) x_s x_t\nonumber\\
                                                                 +2 c_{11} \mu ^2 \cos ^2(2 \Delta ) x_s x_t  -2 c_{23} \mu ^2 \sin (2 \Delta ) y_s x_t+2 c_{32} \mu ^2 \sin (2 \Delta ) x_s y_t\nonumber\\
                                                                 -2 c_{21} \mu ^2 \cos (2 \Delta ) y_s x_t  -2 c_{12} \mu ^2 \cos (2 \Delta ) x_s y_t+2 c_{13} \mu ^2 \sin ^2(2 \Delta ) z_s x_t\nonumber\\
                                                                 +2 c_{31} \mu ^2 \sin ^2(2 \Delta ) x_s z_t +c_{11} \mu ^2 \sin (4 \Delta ) z_s x_t+c_{33} \mu ^2 \sin (4 \Delta ) z_s x_t
                                                                  -c_{11} \mu ^2 \sin (4 \Delta ) x_s z_t\nonumber\\ - c_{33} \mu ^2 \sin (4 \Delta ) x_s z_t 
                                                                  +2 c_{31} \mu ^2 \cos ^2(2 \Delta ) z_s x_t+2 c_{13} \mu ^2 \cos ^2(2 \Delta ) x_s z_t+2 c_{22} \mu ^2 y_s y_t\nonumber\\
                                                                  -2 c_{12} \mu ^2 \sin (2 \Delta ) z_s y_t 
                                                                  +2 c_{21} \mu ^2 \sin (2 \Delta ) y_s z_t-2 c_{32} \mu ^2 \cos (2 \Delta ) z_s y_t-2 c_{23} \mu ^2 \cos (2 \Delta ) y_s z_t \nonumber\\
                                                                  -2 c_{11} \mu ^2 \sin ^2(2 \Delta ) z_s z_t+c_{13} \mu ^2 \sin (4 \Delta ) z_s z_t-c_{31} 
                                                                  \mu ^2 \sin (4 \Delta ) z_s z_t+2 c_{33} \mu ^2 \cos ^2(2 \Delta ) z_s z_t+2 \Big]
                                                                \end{align}

                                                                \end{widetext}

\end{document}